\documentclass[pra,twocolumn,amsmath,amssymb,superscriptaddress,showpacs]{revtex4}
\usepackage{graphics}
\usepackage{epsfig}
\usepackage{bbm}
\usepackage[latin1]{inputen c} 

\newcommand{\be}{\begin{equation}}
\newcommand{\ee}{\end{equation}}
\newcommand{\eea}{\end{eqnarray}}

\newcommand{\exs}[1]{\ensuremath{\langle{#1}\rangle}}

\newcommand{\ket}[1]{\ensuremath{|#1\rangle}}

\newcommand{\bra}[1]{\ensuremath{\langle#1|}}

\newcommand{\kommentar}[1]{}
\newcommand{\trace}{{\rm Tr}}
\newcommand{\twirl}{{\rm \bf P}}
\newcommand{\twirliso}{{\rm \bf P}_{\rm iso}}
\newcommand{\twirlrec}{{\rm \bf Q}}
\newcommand{\twirlbias}{{\rm \bf \tilde P}}
\newcommand{\twirlrecbias}{{\rm \bf \tilde Q}}

\begin{document}
\title{Efficient algorithm for multi-qudit twirling for ensemble quantum computation}
\date{\today}
\begin{abstract}
We present an efficient algorithm for twirling a multi-qudit quantum
state. The algorithm can be used for approximating the twirling
operation in an ensemble of physical systems in which the systems
cannot be individually accessed. It can also be used for computing
the twirled density matrix on a classical computer. The method is
based on a simple non-unitary operation involving a random unitary.
When applying this basic building block iteratively, the mean
squared error of the approximation decays exponentially. In
contrast, when averaging over random unitary matrices the error
decreases only algebraically. We present evidence that the unitaries
in our algorithm can come from a very imperfect random source or can
even be chosen deterministically from a set of cyclically
alternating matrices. Based on these ideas we present a quantum
circuit realizing twirling efficiently.
\end{abstract}

\author{G\'eza T\'oth}
\email{toth@alumni.nd.edu} \affiliation{Research Institute for Solid
State Physics and Optics, Hungarian Academy of Sciences,  P.O. Box
49, H-1525 Budapest, Hungary}
\author{Juan Jos\'e Garc\'{\i}a-Ripoll}
\affiliation{ Max Planck Institute for Quantum Optics,
Hans-Kopfermann-Stra{\ss}e 1, D-85748 Garching, Germany}


\maketitle



\section{Introduction}

Twirling was first introduced for bipartite systems in
Refs.~\cite{BB96,BD96} in the context of entanglement purification
and still appears as part of various quantum information processing
protocols \cite{DL02,BD05,DH05}. For example, in the single party
case, twirling makes possible to obtain the average gate fidelity of
a positive map \cite{EA05}. Later twirling was generalized to
multipartite systems: For a given density matrix $\rho$ the twirled
state is defined as \cite{W89,EW01,EggelingPhD}
\begin{equation}
\twirl \rho := \int_{U\in U(d)}  U^{\otimes N} \rho (U^{\otimes
N})^\dagger dU, \label{integral}
\end{equation}
where $U(d)$ is the group of $d$-dimensional unitary matrices, $N$
is the number of qudits, and $dU$ is the normalized Haar measure
over $U(d).$ For the bipartite case one can also consider twirling
defined as
\begin{equation}
\twirliso \rho := \int_{U\in U(d)}  U \otimes U^* \rho (U \otimes U^*)^\dagger dU,
\end{equation}
where '$^*$' denotes element-wise complex conjugation. States
obtained from $\twirl$ and $\twirliso$ are called Werner states and
isotropic states, respectively. Isotropic states are quite useful in
quantum information processing: They are the maximally entangled
state mixed with white noise. While in this work we focus on
$\twirl,$ our results generalize trivially to the computation of
$\twirliso.$

The importance of twirling in the multipartite case is that it
transforms a general mixed state into a state that can be
characterized with only a few parameters \cite{W89,CK06,remark1}.
Entanglement of formation is known for bipartite isotropic states
\cite{TV00,FL06} and necessary and sufficient conditions for the
entanglement of tripartite Werner states are also known \cite{EW01}.
Therefore, since twirling cannot increase any entanglement monotone,
if we can experimentally twirl a state, we can simplify the
estimation of its entanglement properties.

Moreover, twirling also appears in various calculations in quantum
information science (e.g., it is used to define a family of quantum
states in Ref.~\cite{TA06}). Integrals over $U(d)$, similar to
twirling, appear in many areas of physics \cite{integU}. In
particular, the computation of integrals of the form
\begin{equation}
\int_{U\in U(d)}
U_{i_1j_1}U_{i_2j_2}...U_{i_mj_m}(U_{k_1l_1}U_{k_2l_2}...U_{k_nl_n})^*
dU, \label{UUU}
\end{equation}
are needed. Such integrals for the $m=n$ case can straightforwardly
be obtained from twirling appropriately chosen density matrices.
Twirling is also closely related to unitary $t$-designs which have
raised interest recently \cite{DC06,GA06}.

It seems straightforward to implement twirling: One has to apply a
random multilateral unitary rotation to each copy of a state, and
then average over the ensemble. However, problems quickly arise when
considering practical implementations. Applying different random
rotations to different systems of the ensemble requires that we are
able to access the systems individually. In practice, very often
this also means temporal avaraging \cite{AX05}: We repeat many times
the following two steps: (i) Generate the quantum state and (ii)
apply a random rotation. The disadvantage of this approach is that
the execution time is proportional to the number of systems in the
ensemble. Moreover, in many physical realizations of quantum
computing, e.g., in a Nuclear Magnetic Resonance (NMR) quantum
computer, this approach cannot be used since the systems cannot be
individually accessed. From the numerical point of view, the problem
is that averaging over the randomly rotated matrices is a very
inefficient way for calculating the integral in
Eq.~(\ref{integral}).

Another approach is using group theory to replace the integral
(\ref{integral}) with a sum over a finite number of rotated density
matrices \cite{AX05_cite}. This works for small systems and, for
example, for $N=2$ and $d=2$ we need to employ $12$ such matrices
\cite{BD96,A97}. However, the number of unitaries needed increases
rapidly with $N$ and $d$, making the implementation of twirling for
large systems difficult this way \cite{DC06}. Clearly, this approach
does not seem to fit ensemble quantum computing. From the point of
view of a realization on a digital computer, there is the added
complexity of computing the required unitaries when compared to
averaging over random unitaries.

Numerically, there is another approach for replacing the integration
with a discrete sum. The idea is that twirling transforms any quantum
state into a $U^{\otimes N}$ invariant state \cite{W89}. The density
matrix of such a state can be written as $\rho=\sum_k \exs{R_k}_\rho
R_k$ where the $R_k$ basis operators are obtained from orthogonalizing
the $N!$ permutation operators \cite{EW01,EggelingPhD}. Since twirling
does not change the expectation values of $R_k,$ we can use these
values to reconstruct the final Werner state, $\twirl\rho$, on a
classical computer. However, once more, while this approach is
feasible for small systems (For $N=2$ and $N=3$ qubits there are $2$
and, respectively, $5$ such matrices \cite{EW01,EggelingPhD}), for
large $N$ the number of permutation matrices increases dramatically.

\begin{figure}
\centerline{\epsfxsize2.5in\epsffile{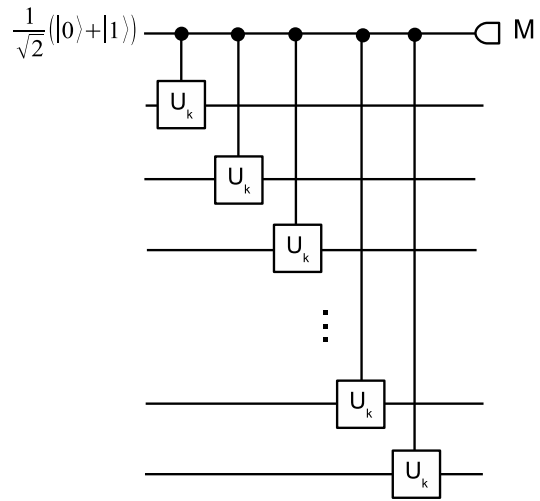}}
\vskip
0.5cm \caption{Twirling can be realized by the repeated application
of this basic building block where $U_k$ is a random unitary
generated for the $k$th iteration or a unitary chosen from a
cyclically alternating set of unitaries. 'M' represents measurement
in the computational basis.} \label{circuit}
\end{figure}

In this paper we show that a multi-qudit state can be approximately
twirled by iterating the single non-unitary operation
\begin{equation}
\rho_{k+1}=\frac{1}{2}\left[
\rho_k+U_k^{\otimes N}\rho_k(U_k^{\otimes N})^\dagger\right],
\end{equation}
where $\rho_k$ are density matrices and
$U_k$ are
random unitaries. Using this building block, the error of our
approximation decays exponentially with the number of iterations.
The exponent of this decay depends neither on the number of qudits
nor on their dimension.  In contrast, when
approximate twirling is realized by averaging density matrices
obtained from multi-lateral random rotations, the convergence is
algebraic. We show evidence that the unitary matrices applied can
come from a highly imperfect source and also demonstrate through
examples that the random unitaries can be replaced by a set of
cyclically alternating unitaries, while preserving the exponential
convergence. Finally, based on the previous ideas, we present a
quantum circuit for twirling by means of controlled unitary gates
(see Fig.~\ref{circuit}).

Experimental implementation of this operation looks feasible in many
physical systems. It is important to stress that, when applied to an
ensemble of many systems, our method does not need individual access
to the individual systems. We show that for a given set of
cyclically alternating unitaries it is possible to obtain general
statements for the convergence which are valid for all density
matrices. This makes it possible to design algorithms tailored for
the operators available in a given physical system. Thus twirling
can be one of the quantum algorithms which are especially fitting
for realization on a quantum computer. On the other hand, when
realizing our algorithm on a classical computer, the programming and
computational effort is extremely small.

Our paper is organized as follows. In Sec.~\ref{usual_method} we
discuss the usual way twirling is computed on a quantum or a
classical computer. In Sec.~\ref{recursive_method} we present our
proposal, together also with an analysis of the convergence of the
approach. In Sec.~\ref{errors} we show that our method is quite
robust against the imperfections of the random number generator.  In
Sec.~\ref{deterministic} we show that instead of random unitaries,
cyclically alternating operators can also be efficiently used for
twirling. In Sec.~\ref{largedim} we discuss the case of large
dimensions. In Sec.~\ref{Experiments} we explain how to use our
ideas for experiments. In Sec.~\ref{Numerics} we show how to
generalize our method for the numerical integration of useful
formulas over the unitary group. Finally, in Sec.~\ref{disc} we
discuss connections of our research to existing work.

\section{Straightforward numerical integration}
\label{usual_method}

$\twirl \rho$ can be approximated by an average of a finite number of
randomly rotated density matrices
\begin{equation}
\label{twirl_M} \twirl_M \rho :=
\frac{1}{M}\left(\rho+\sum_{k=1}^{M-1}U_k^{\otimes N} \rho
(U_k^{\otimes N})^\dagger\right).
\end{equation}
Here $M$ denotes the number of terms and we assume that the unitaries
$\{U_k\}$ are distributed uniformly in $U(d)$ according to
the Haar measure. In this section we examine how well $\twirl_M \rho$
converges to $\twirl \rho$ for increasing $M.$

Since we use random matrices, we will obtain a different state for
each realization of $\twirl_M \rho$. To analyze the error, we
introduce an expectation value or average over the different choices
for $U_k$ as \cite{exA}
\begin{equation}
\exs{A}:=\int A dU_1 dU_2 dU_3 ...
\end{equation}
Using this average, we can analyze how fast $\twirl_M \rho$
converges to $\twirl \rho$ for increasing number of unitaries.
Simple calculations show that the average error of a particular
initial state, $\rho$, decreases algebraically as $M^{-1}$
\begin{eqnarray}
&&\exs{\|\twirl_M\rho-\twirl\rho\|^2}\nonumber\\
&&\;\;\;\;\;\;\;\;\;\;\;\; =
\exs{\|\twirl_{M}\rho\|^2}+\|\twirl\rho\|^2-
2\trace(\exs{\twirl_M\rho}\twirl\rho)\nonumber\\
&&\;\;\;\;\;\;\;\;\;\;\;\;=\exs{\|\twirl_{M}\rho\|^2}-\|\twirl\rho\|^2\nonumber\\
&&\;\;\;\;\;\;\;\;\;\;\;\;=\frac{1}{M}\left(\|\rho\|^2-\|\twirl\rho\|^2\right),
\label{error_usual}
\end{eqnarray}
where $\|A\|^2:=\trace(A^\dagger A)$ is the Hilbert-Schmidt norm. In
the derivation we used that
$\exs{\twirl_M\rho}=(1/M)\rho+[(M-1)/M]\twirl \rho$ and
$\trace(\rho\twirl\rho)=\trace[(\twirl\rho)^2].$

While computing the error for a given state is illuminating, it is
more useful to characterize the convergence of $\twirl_M$ in a
manner that is independent of the initial state. For that, first we
will show how to define a matrix describing the action of a linear
superoperator and will define a measure of distance between
superoperators. Then, we will determine the matrices describing the
action of $\twirl$ and $\twirl_M,$ and will compute the norm of
their difference.

Density matrices are vectors in a Hilbert space of complex matrices
with the scalar product $\langle \rho,\rho'\rangle :=
\mathrm{Tr}(\rho\rho')$. Thus it is convenient to switch from matrix
notation
\begin{equation}
\rho=\sum_{kl} \rho_{kl} \ket{k}\bra{l}\
\end{equation}
and treat the matrices as vectors defined by \cite{otimes}
\begin{equation}
\vec{\rho}=\sum_{kl} \rho_{kl} \ket{l} \otimes \ket{k}.
\end{equation}
That is, $\vec{\rho}$ is obtained from $\rho$ by joining its columns
consecutively into a column vector. We can use the vector form for any
Hermitian operator $A$, not only for density matrices. Then the
expectation value of $A$ can be written as
\begin{equation}
\trace(A\rho)=(\vec{A})^\dagger\vec{\rho}.
\label{vecA}
\end{equation}
Any physically allowed transformation of the density matrix is a
linear positive map and it can be written as a matrix acting on $\vec{\rho}$
\begin{equation}
\vec{\rho}\;'=S \vec{\rho}.
\end{equation}
Matrix $S$ describes the transformation realized by the
superoperator. Both the vectors $\vec\rho$, $\vec{\rho}\;'$, and the
matrix $S$ have to satisfy constraints to ensure the Hermiticity and
the positivity of the density matrices. The distance between
superoperators can be measured in the form of Hilbert-Schmidt norm
of their difference \cite{norm}
\begin{equation}
\Vert S - \tilde S\Vert^2 := \mathrm{Tr}[(S-\tilde S)(S-\tilde
S)^\dagger]. \end{equation}

In this formalism, the superoperators describing the action of
$\twirl$ and $\twirl_M$ are, respectively,
\begin{subequations}
\label{SP}
\begin{eqnarray}
S_{\twirl}&=&\int_{U\in U(d)}  (U^{\otimes N})^{*} \otimes
U^{\otimes
N} dU, \\
S_{\twirl M}&=&\frac{1}{M}\left(\openone_{d}^{\otimes 2N}
+\sum_{k=1}^{M-1}(U_k^{\otimes N})^{*} \otimes U_k^{\otimes
N}\right),
\end{eqnarray}
\end{subequations}
where $\openone_{d}$ denotes a $d\times d$ unit matrix.
Based on Eq.~(\ref{SP}a), it is easy to see that
\begin{equation}
S_{\twirl}S_{\twirl}=S_{\twirl}=S_{\twirl}^\dagger.
\label{Sprop}
\end{equation}
Using these, straightforward calculation shows that
\begin{eqnarray}
&&\exs{\|S_{\twirl M}-S_{\twirl}\|^2}\nonumber\\
&&\;\;\;\;\;\;\;\;\;\;\;=\exs{\|S_{\twirl
M}\|^2}+ \|S_{\twirl}\|^2-2\trace(S_{\twirl}\exs{S_{\twirl M}})\nonumber\\
&&\;\;\;\;\;\;\;\;\;\;\;=\frac{1}{M}\left(\|\openone_{d}^{\otimes
2N}\|^2-\|S_{\twirl}\|^2\right). \label{error_usualS}
\end{eqnarray}
Thus the error in the superoperator formalism decays algebraically
with increasing number of steps, $M$, irrespective of the initial
state (see Appendix \ref{SecSP} for details).

\section{Twirling using a recursive formula}
\label{recursive_method}

In order to decrease the error of the result, rather than doubling
the number of terms in the summation and computing $\twirl_{2M}$, we
can apply twice the averaging operation with $(M-1)$ unitaries and
calculate $\twirl_{M}\twirl_{M}\rho$. In this section we show that,
even though in both cases $\sim 2M$ random unitaries are needed, the
error of the second method is much smaller.

Let us write out the result after two twirlings explicitly:
\begin{eqnarray}
\twirl_{M}\twirl_{M}\rho&=&\frac{1}{M^2}
\sum_{k=1}^{M-1}\sum_{l=1}^{M-1}\rho+U_k^{\otimes N} \rho
[(U_k^{\otimes N}]^\dagger\nonumber\\
&+&U_{M-1+l}^{\otimes N} \rho [(U_{M-1+l}^{\otimes
N}]^\dagger\nonumber\\&+&(U_{M-1+l}U_k)^{\otimes N} \rho
[(U_{M-1+l}U_k)^{\otimes N}]^\dagger,\nonumber\\ \label{twot}
\end{eqnarray}
where $\{U_k\}_{k=1}^{M-1}$ and $\{U_k\}_{k=M}^{2M-2}$ are the
random unitaries chosen for the first and second twirling,
respectively. Eq.~(\ref{twot}) is the average of $M^2-1$ rotated
density matrices and the original matrix. We have the same number of
terms when computing $\twirl_{M^2}\rho.$ However, the $M^2$
unitaries are not independent thus we might expect that the error
for $\twirl_{M}\twirl_{M}\rho$ is larger than that for
$\twirl_{M^2}\rho.$

Let us now consider repeated applications of $\twirl_m.$ For
simplicity we will first focus on the $m=2$ case, leaving the $m>2$
case for later. After $M$ iterations, the outcome is
\begin{equation}
\twirlrec_{M}\rho:=\twirl_2 \twirl_2 ... \twirl_2 \rho=
\left(\prod_{k=1}^M \twirl_2\right) \rho. \label{twirlrec}
\end{equation}
Using the definition Eq.~(\ref{twirlrec}) we can write the recursive
formula
\begin{eqnarray}
\twirlrec_{M} \rho=\frac{1}{2}\left[ \twirlrec_{M-1}\rho +
U_M^{\otimes N} \ (\twirlrec_{M-1}\rho) (U_M^{\otimes N})^\dagger
\right], \label{rec_twirlrec}
\end{eqnarray}
where again $U_M$ is a random unitary. As before, we measure the
convergence of this operator by the average error in the
Hilbert-Schmidt norm
\begin{eqnarray}
\exs{\|\twirlrec_M\rho-\twirl\rho\|^2}
&=&\exs{\|\twirlrec_{M}\rho\|^2}-\|\twirl\rho\|^2.
\label{tildetwirl}
\end{eqnarray}
For computing the error as a function of $M,$ we need the
$M$-dependence of the $\exs{\|\twirlrec_{M}\rho\|^2}$ term on the
right hand side of Eq.~(\ref{tildetwirl}). For that first we express
$\exs{\|\twirlrec_{M}\rho\|^2}$ with
$\exs{\|\twirlrec_{M-1}\rho\|^2}$
\begin{eqnarray}
\exs{\|\twirlrec_{M}\rho\|^2}&=&\frac{1}{2}\big[
\exs{\|\twirlrec_{M-1}\rho\|^2}\nonumber\\&+&\exs{\|\twirlrec_{M-1}\rho
U_M^{\otimes N}
\ (\twirlrec_{M-1}\rho) (U_M^{\otimes N})^\dagger\|^2}\big]\nonumber\\
&=&\frac{1}{2}\left(
\exs{\|\twirlrec_{M-1}\rho\|^2}+\|\twirl\rho\|^2\right) .
\label{tildetwirl2}
\end{eqnarray} Then, from
Eq.~(\ref{tildetwirl2}) the $M$-dependence of
$\exs{\|\twirlrec_M\rho\|^2}$ can be obtained as
\begin{equation}
\exs{\|\twirlrec_M\rho\|^2}=\|\rho\|^2+
\left(\|\twirl\rho\|^2-\|\rho^2\|\right)(1-2^{-M}).
\label{tildetwirl2b}
\end{equation}
Substituting Eq.~(\ref{tildetwirl2b}) into Eq.~(\ref{tildetwirl}) we
obtain
\begin{equation}
\exs{\|\twirlrec_M\rho-\twirl\rho\|^2}=
\left(\|\rho\|^2-\|\twirl\rho\|^2\right)2^{-M}.
\label{error_recursive_rho}
\end{equation}
That is, the squared error decays exponentially with $M,$ while according to
Eq.~(\ref{error_usual}) the decay was $\propto M^{-1}$ for the
method described in Sec.~\ref{usual_method}. Note that computing
$\twirlrec_M\rho$ and $\twirl_M \rho$ need the generation of $M$ and
$M-1$ random unitaries, respectively. Thus the computational effort
is roughly the same for the two cases.

One can repeat this calculation in the superoperator picture. the
The definition of $S_{\twirlrec M}$ based on
Eq.~(\ref{rec_twirlrec}) is
\begin{equation}
S_{\twirlrec M} =\frac{1}{2}\left\{ S_{\twirlrec (M-1)} +
[(U_M^{\otimes N})^* \otimes  U_M^{\otimes N}]\ S_{\twirlrec (M-1)}
\right\}. \label{rec_twirlrec_S}
\end{equation}
For the error of the approximation we obtain
\begin{eqnarray}
\exs{\|S_{\twirlrec M}-S_{\twirl}\|^2}&=&
\exs{\|S_{\twirlrec M}\|^2}+\|S_{\twirl}\|^2\nonumber\\&-&
\exs{\trace(S_{\twirlrec M}^\dagger S_{\twirl})}-\exs{\trace(S_{\twirlrec M}S_{\twirl})}\nonumber\\
&=&\exs{\|S_{\twirlrec M}\|^2}-\|S_{\twirl}\|^2.
\label{errorS}
\end{eqnarray}
For obtaining the $M$-dependence of the error, we need the
$M$-dependence of $\|S_{\twirlrec M}\|^2$. This is obtained in two
steps. First, we use Eq.~(\ref{rec_twirlrec_S}) to find a recursive
relation for $\exs{\|S_{\twirlrec M}\|^2}$
\begin{equation}
\exs{\|S_{\twirlrec M} \|^2}=\frac{1}{2}\left(
\exs{\|S_{\twirlrec (M-1)}\|^2}+
\|S_{\twirl}\|^2\right), \label{tildetwirl2_S}
\end{equation}
and then we obtain $\exs{\|S_{\twirlrec M}\|^2}$ without using recursion as
\begin{equation}
\exs{\|S_{\twirlrec M}\|^2}=\|\openone_{d}^{\otimes 2N}\|^2+
\left(\|S_{\twirl}\|^2-\|\openone_{d}^{\otimes 2N}\|^2\right)(1-2^{-M}).
\label{tildetwirl2b_S}
\end{equation}
Combining Eq.~(\ref{tildetwirl2b_S}) and (\ref{errorS}) we obtain
\begin{equation}
\exs{\|S_{\twirlrec M}-S_{\twirl}\|^2}=
\left(\|\openone_{d}^{\otimes 2N}\|^2-\|S_{\twirl}\|^2\right)2^{-M},
\label{error_recursive}
\end{equation}
thus the error of the superoperator decays exponentially.

Let us now consider combining twirl operations $\twirl_K$ with more
than two unitaries. With that aim we define
\begin{equation}
\twirlrec_{K,M}:=(\twirl_K)^M.
\end{equation}
On the one hand, when computing the average error for this operator we
obtain formulas similar to Eq.~(\ref{error_recursive_rho}) and
Eq.~(\ref{error_recursive}), but with a faster decay $\propto K^{-M}$
vs. the original $\propto 2^{-M}.$ On the other hand, the number of
random unitaries required increases, as it is now $K-1$ per iteration
step. Based on these we can write
 the dynamics of the mean squared error as a function of
the number of unitaries $N_U$ as
\begin{equation}
\langle \Vert \twirlrec_{K,M} - \twirl\Vert^2\rangle \propto
\exp\left(-\frac{\ln K}{K-1}N_U\right).
\end{equation}
Hence one can see that for a given number of unitaries, the smallest error
is achieved for $K=2.$ For many experiments, this is also a good
reasoning since the experimental effort is very often measured in
$N_U.$ Thus we will consider the $K=2$ case in the rest of the paper.

We have verified numerically the previous results
\cite{QUBIT4MATLAB}. For this we focused on the three-qubit case,
for which we can compute the twirl operation exactly using the
techniques mentioned in the introduction [See Appendix \ref{SecSP}].
In Fig.~\ref{error_timedep} we plot the error $\exs{\|S_{\twirl
M}-S_{\twirl}\|^2}$ averaged over $10000$ trajectories. As the
figure shows, the simulation results perfectly fit the exponential
decay of the error calculated theoretically (\ref{error_recursive}).

\begin{figure}
\vskip 0.5cm \centerline{ \epsfxsize 2.8in
\epsffile{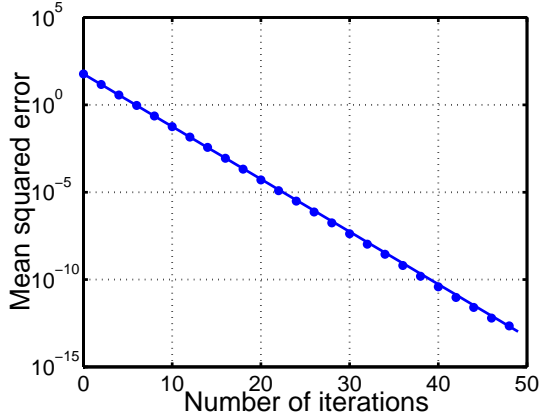}}
\caption{Mean squared error for the recursive method with random
  matrices when applied on three-qubit states. We plot (dotted) the average
  over 10000 realizations and (solid line) the theoretical prediction
  computed from Eq.~(\ref{error_recursive}).
For better visibility, the error is shown only for every second
iteration. } \label{error_timedep}
\end{figure}

\section{Sensitivity to the imperfections of random number generation}
\label{errors}

In this section we examine what happens if our random number
generator does not work perfectly and the random unitaries are not
uniformly distributed over $U(d).$ An imperfect random unitary
generator can be characterized by the distribution $f(U)$ describing
the probability density for getting $U.$ We will show that if
$\inf_U f(U) >0$ then our algorithm still converges to the twirled
state and the error decays exponentially.

Let us use a simple model for our faulty distribution in which with
probability $p_g$ the unitary is drawn according to the probability
distribution $g(U)$ while with probability $(1-p_g)$ it is drawn
according to the uniform distribution. The corresponding
distribution function $f(U)$ is
\begin{equation}
f(U):=p_gg(U)+(1-p_g), \label{fUUU}
\end{equation}
where we used that $\int_{U\in U(d)} dU=1.$ Expectation values over
this probability distribution are computed as
\begin{eqnarray}
&&\exs{A}_f:=\int_{U\in U(d)} A f(U_1) f(U_2) f(U_3) ... dU_1 dU_2
dU_3 ...\nonumber
\end{eqnarray}

Let us now examine how the usual method described in
Sec.~\ref{usual_method} is affected by such an error of the random
number generator. We will define by $\twirlbias_M$ the equivalent of
Eq.~(\ref{twirl_M}) with our biased probability distribution. The mean
value of the density matrix obtained from such twirling is
\begin{eqnarray}
&&\exs{\twirlbias_M \rho}_f=
\frac{1}{M}\bigg(\rho+(1-p_g)(M-1)\twirl \rho
+\nonumber\\
&&\;\;\;\;\;\;+p_g(M-1)  \int dW g(W) W^{\otimes N} \rho (W^{\otimes
N})^\dagger\bigg).\nonumber
\end{eqnarray}
Here $\int dW$ is an integral over the unitary group $U(d).$ Taking
the limit $M\rightarrow\infty$ one obtains
\begin{eqnarray}\exs{\twirlbias_M\rho}_f\rightarrow (1-p_g)\twirl\rho +
p_g \int dW g(W) W^{\otimes N} \rho (W^{\otimes
    N})^\dagger.\nonumber
\end{eqnarray}
Thus the expectation value of the operator does not converge to
$\twirl \rho.$

On the contrary, when $\twirlbias_2$ is applied $M$ times, the state
of the system still converges to the twirled state and the error
decays exponentially with $M.$ To show this, let us denote the
operation above by $\twirlrecbias_M.$ As before, we measure the
convergence of this operator by the average error in the
Hilbert-Schmidt norm
\begin{equation}
\exs{\|\twirlrecbias_M\rho-\twirl\rho\|^2}_f=
\exs{\|\twirlrecbias_{M}\rho\|^2}_f-\|\twirl\rho\|^2.
\label{tildetwirlbias}
\end{equation}
Hence straightforward algebra yields
\begin{equation}
\exs{\|\twirlrecbias_{M}\rho\|^2}_f\le\tfrac{1+p_g^2}{2}
\exs{\|\twirlrecbias_{M-1}\rho\|^2}_f+\tfrac{1-p_g^2}{2}
\|\twirl\rho\|^2. \label{tildetwirl2_bias}
\end{equation}
For obtaining the upper bound
in Eq.~(\ref{tildetwirl2_bias})
we used
\begin{eqnarray}
\trace\left[\twirlrecbias_{M-1}\rho U^{\otimes N} \
(\twirlrecbias_{M-1}\rho) (U^{\otimes N})^\dagger\right]&\le&
\|\twirlrecbias_{M-1}\rho\|^2,\nonumber\\
\bigg\|\int dW g(W) W^{\otimes N} \ (\twirlrecbias_{M-1}\rho)
(W^{\otimes N})^\dagger \bigg\|^2&\le&
\|\twirlrecbias_{M-1}\rho\|^2,\nonumber\\
\end{eqnarray}
where $U$ is a unitary matrix. From Eq.~(\ref{tildetwirl2_bias}) the
$M$-dependence of $\exs{\|\twirlrecbias_M\rho\|^2}_f$ can be deduced
as
\begin{eqnarray}
&&\exs{\|\twirlrecbias_M\rho\|^2}_f\le\|\rho\|^2\nonumber\\&&\;\;\;\;\;\;\;\;
+\left(\|\twirl\rho\|^2-\|\rho^2\|\right)
\left[1-\left(\frac{2}{1+p_g^2}\right)^{-M}\right].\nonumber\\
\label{tildetwirl2b_bias}
\end{eqnarray}
Substituting Eq.~(\ref{tildetwirl2b_bias}) into
Eq.~(\ref{tildetwirlbias}) we obtain
\begin{equation}
\exs{\|\twirlrecbias_M\rho-\twirl\rho\|^2}_f\le
\left(\|\rho\|^2-\|\twirl\rho\|^2\right)\left(\frac{2}{1+p_g^2}\right)^{-M}.
\label{error_recursive_rho_bias}
\end{equation}
The mean square error of the superoperator corresponding to
$\twirlrecbias_M$ also decays as $\propto[2/(1+\textstyle
p_g^2)]^{-M}.$ Thus we have convergence if $p_g<1,$ i.e., if the
uniform distribution has a non-zero weight in Eq.~(\ref{fUUU}). For
functions $f(U)$ that satisfy
\begin{equation}\inf_U f(U) \label{sc}
>0,\end{equation}
it is always possible to find a decomposition of the type
Eq.~(\ref{fUUU}) such that $p_g=1-\inf_U f(U).$ Thus for such
probability distribution functions our algorithm converges and the
error decays exponentially.

Finally, the sufficient condition for the convergence of our method
Eq.~(\ref{sc}) can also be formulated for the case when $f(U)$ is of
the form
\begin{equation}
f(U)=f_{r}(U)+\sum_k c_k\delta(U-V_k),
\end{equation}
where $f_r:U(d)\mapsto\mathbb{R},$ $c_k\ge 0$ are constants,
$\delta$ is the Dirac delta function and $V_k$ are unitaries. In
this case the algorithm converges if
\begin{equation}
\inf_{e(U)} \int_{U\in U(d)} f(U)e(U) dU>0,
\end{equation}
where for the function $e(U)$ we require that $e(U)\ge 0$ and
$\int_{U\in U(d)} e(U)dU=1.$

\section{Deterministic Twirling with few unitaries}
\label{deterministic}

The example shown in Sec.~\ref{errors} demonstrated that even if the
random unitaries used in our algorithm come from a very imperfect
source, the algorithm may still converge. In this section we will
examine what happens if these unitaries are not random but they are
chosen deterministically from a small set such that they are
cyclically alternating.

Let us consider the two-qubit case. Common sense tells us that we need
at least two unitaries since, two unitaries are able to generate the
elements of $U(2)$ we need for twirling \cite{SU}. We will see that
two unitaries are sufficient.

\begin{figure}
\vskip 0.5cm \centerline{\epsfxsize 2.8in
\epsffile{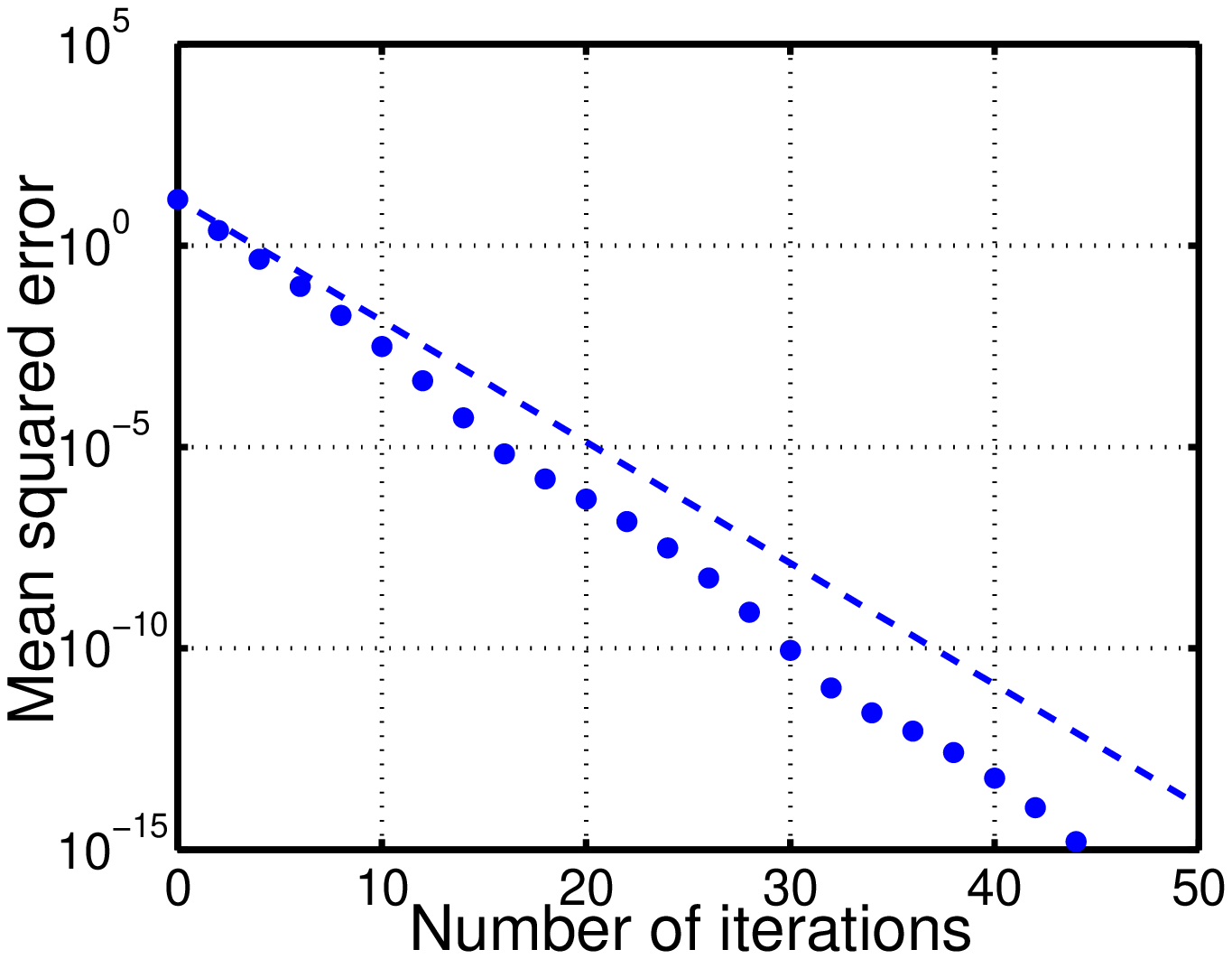} } \centerline{(a)}
\vskip 0.5cm \centerline{\epsfxsize 2.8in
\epsffile{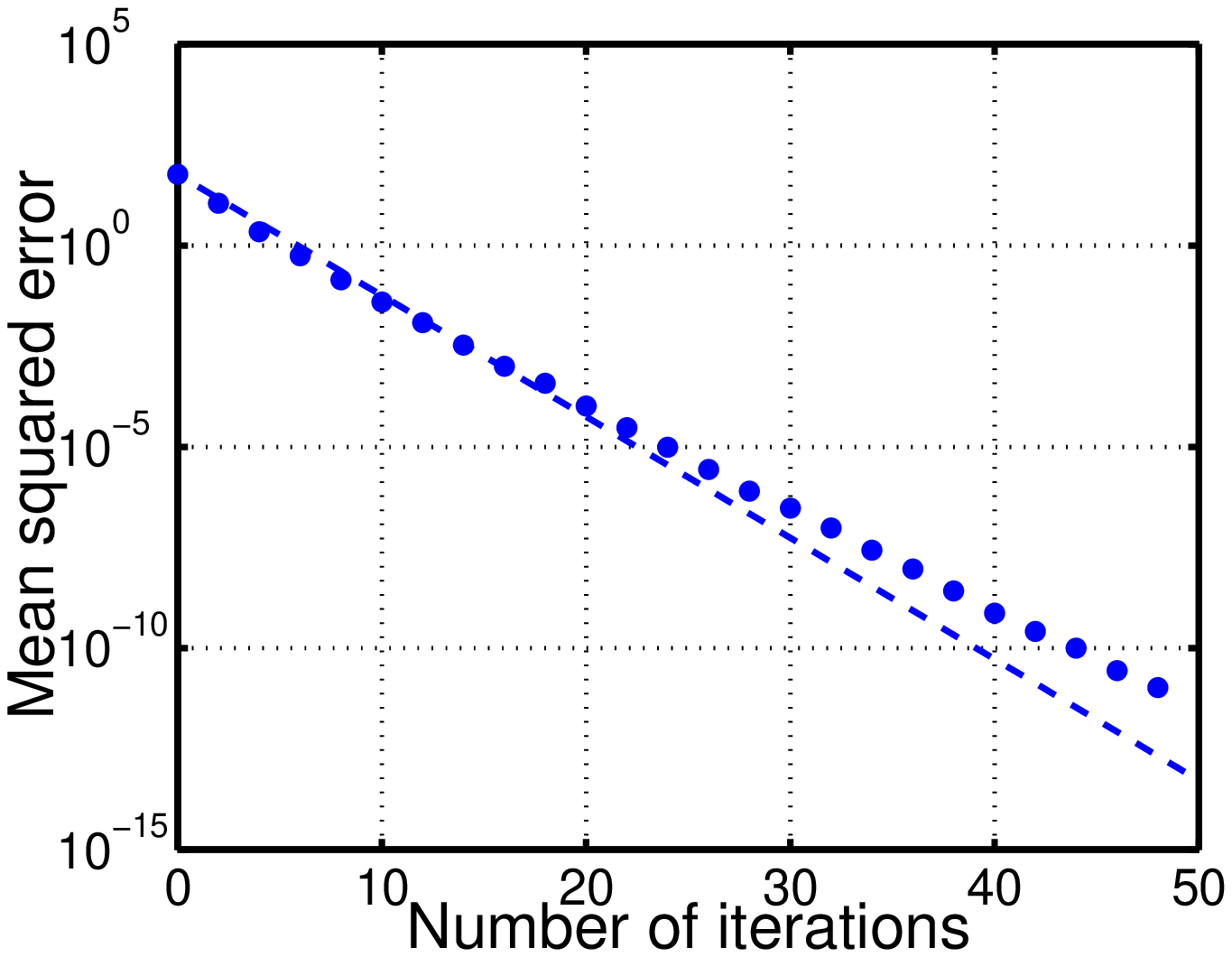} } \centerline{(b)}
\vskip 0.5cm \caption{ Time dependence of the error (a) for two
qubits and (b) for three qubits for the deterministic method using
two and three unitaries, respectively. For better visibility, the
error is shown only for every second iteration. Dashed line
indicates the error for the method using random matrices, given in
Eq.~(\ref{error_recursive}). } \label{error_timedep_deterministic}
\end{figure}

Let us choose the two unitaries as
\begin{eqnarray}
U_x&:=&e^{ic\sigma_x},\nonumber\\
U_z&:=&e^{ic\sigma_z},
\end{eqnarray}
where $\sigma_{x/z}$ are Pauli spin matrices and $c$ is a constant.
Now we can use the
method described at the end of Sec.~\ref{recursive_method} to
compute the dependence of the superoperator on the number of
iterations. We look for the $c$ for which the
decay of the error is the fastest.
Through numerical optimisation we find that
the error of the superoperator is the smallest after $50$ iterations for
$c=1.0894.$ Fig.~\ref{error_timedep_deterministic}(a) shows the
results of our numerical calculations for two qubits with this value for $c.$
The dashed line shows the square of the error for the
recursive method using random unitaries described in
Sec.~\ref{recursive_method}.
Note that the error for the deterministic method, denoted by disks,
decays faster than for the random method.

Let us see a three-qubit example with three cyclically alternating
unitaries
\begin{eqnarray}
U_x&:=&e^{i2\pi/3\sigma_x},\nonumber\\
U_y&:=&e^{i2\pi/5\sigma_y},\nonumber\\
U_z&:=&e^{i2\pi/3\sigma_z}.
\end{eqnarray}
Fig.~\ref{error_timedep_deterministic}(b) shows the results of our
numerical calculations. Now the error decays somewhat more slowly
than in the case of the random method.

The advantage of our approach is that by studying the superoperator,
we can make general statements, independent from the initial state,
about the algorithm. Thus without a thorough group-theoretical study
we can show that the error decays exponentially with $M$ and the
recursive algorithm with the given alternating unitaries can be used
for the efficient twirling of two or three qubits, respectively.
Similar calculations can be carried out for several qubits trying
out other unitaries or higher dimensions can also be investigated.
These calculations can always consider the gates easily available in
an experimental implementation.

\section{Twirling for qudits with large dimensions}
\label{largedim}

In the literature there was a considerable effort to find a method
for two-qudit twirling which can be realized with relatively few
quantum gates even if the dimension of qudits is large. In this
section we show through examples that the number of quantum gates
necessary for two-qudit twirling with our algorithm seems to scale
better with the dimension of the qudits than for the algorithm
generating a $d$-dimensional random unitary uniformly distributed
according to the Haar measure.

A method for generating a random unitary of large dimensions was
presented in Ref.~\cite{EW03}. The system was considered to be a
multi-qubit system which fits well for many physical realizations.
The algorithm presented has two steps: (i) Single-qubit random
unitaries act on the individual qubits. (ii) A nearest-neighbor
Ising interaction acts on the one-dimensional array of qubits. These
two steps must be repeated several times. It was found that the
number of gates necessary for generating a random unitary this way
scales exponentially with the number of qubits $n.$

In Ref.~\cite{DL02} it was shown that two-qudit twirling over
$U(2^n)$ gives the same result as two-qudit twirling over the
Clifford group. This makes efficient twirling possible since the
number of gates needed for generating a random Clifford group
element scales polynomially with $n.$ The results of
Ref.~\cite{DL02} were extended to unitary $2$-designs in
Ref.~\cite{DC06}.

Let us now examine whether it is possible to find an efficient way
to realize our algorithm for $d>2.$ We also consider the $d=2^n$
case. We look for a simple way for generating an imperfect random
unitary such that it still can be used for two-qudit twirling. In
particular, we would like that the error does not decay slower than
when using unitaries which are uniformly distributed according to
the Haar measure.

We use a slight modification of the algorithm presented in
Ref.~\cite{EW03}. Our random $n$-qubit unitary is generated by
applying first different random unitaries for each qubit, then
making the system evolve under an Ising Hamiltonian with
nearest-neighbor interaction realizing
\begin{equation}
U_{\rm Ising}:=\exp\left(i\alpha\sum_k
\sigma_z^{(k)}\sigma_z^{(k+1)}\right),
\end{equation}
where $\alpha$ is a constant and we consider a periodic boundary
condition. Thus we use a single iteration of the method presented in
Ref.~\cite{EW03}. The gate requirements increase linearly with $n$
for such an algorithm.

Next we show simulations with the density matrix rather than
simulations with the superoperator. The reason is that the size of
the superoperator is $16^n\times16^n$ which would make it possible
to consider only small systems. We calculate the dynamics obtained
from our algorithm for several random density matrices which have a
uniform distribution according to the Hilbert-Schmidt measure
\cite{ZS01}. In order to compare trajectories corresponding to
different density matrices, we compute the normalized error
\begin{equation}
E_{\rm norm}:= \frac{\exs{\|\twirlrec_M\rho-\twirl\rho\|^2}}{
\left(\|\rho\|^2-\|\twirl\rho\|^2\right)}. \label{normalized}
\end{equation}
It follows from Eq.~(\ref{error_recursive_rho}) that for the method
using random matrices uniformly distributed over $U(d)$ we have
$E_{\rm norm}=2^{-M}.$

\begin{figure}
\vskip 0.5cm \centerline{\epsfxsize 2.8in
\epsffile{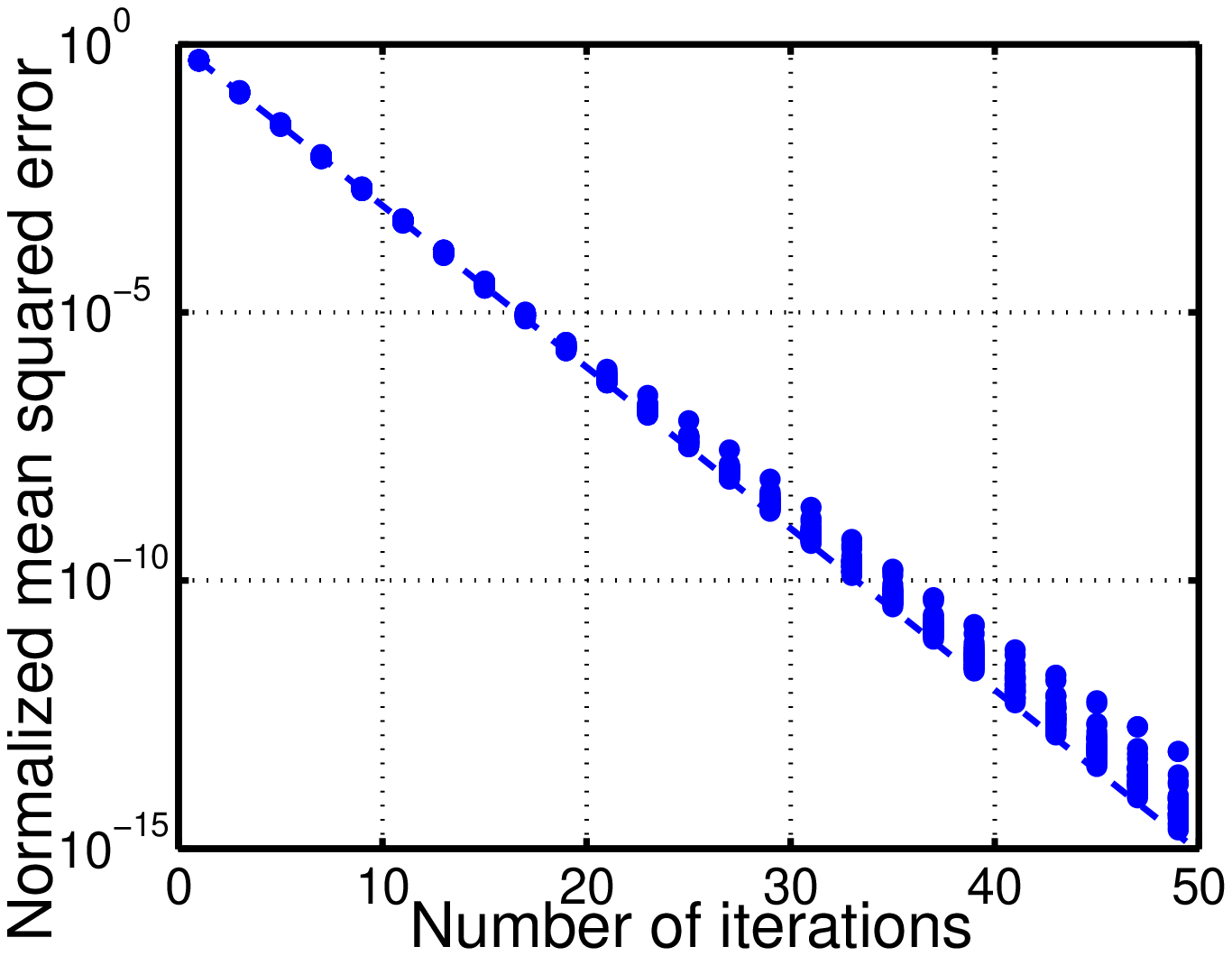} } \centerline{(a)} \vskip 0.5cm
\centerline{\epsfxsize 2.8in \epsffile{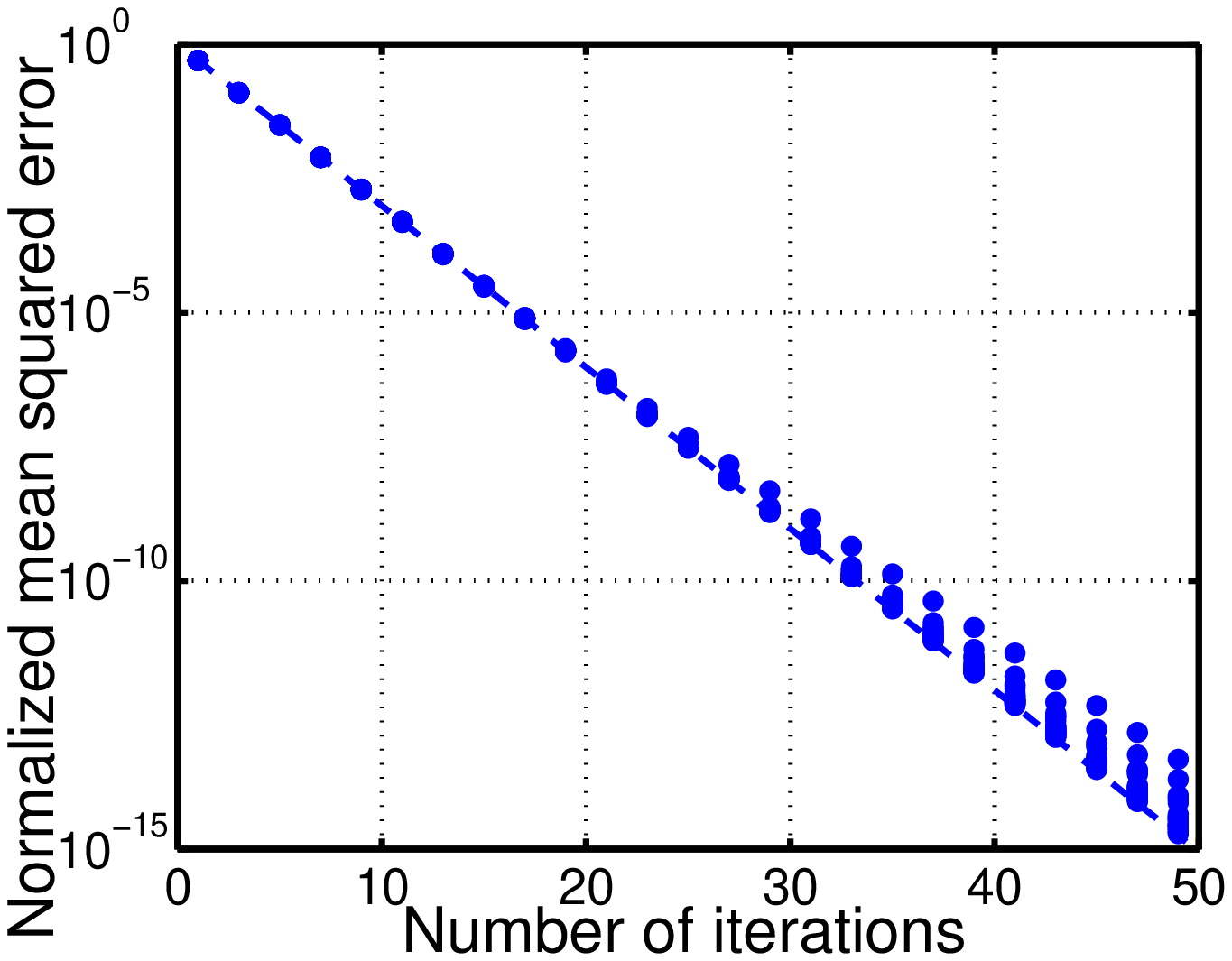} }
\centerline{(b)} \vskip 0.5cm \caption{Time dependence of the
normalized error of the density matrix for bipartite twirling for
(a) $d=8$ and (b) $d=16.$ In both cases 25 realizations are shown.
See text for the algorithm used for generating $d$-dimensional
unitaries. For better visibility, the error is shown only for every
second iteration. Dashed line indicates the error for the method
using random matrices uniformly distributed over $U(d).$}
\label{error_timedep_larged}
\end{figure}

Fig.~\ref{error_timedep_larged} shows the results of our
calculations for $d=2^3$ and $d=2^4$. We used $\alpha=1.10$ and
$1.03$, respectively. We find that the error decays almost as in the
case of using random unitaries uniformly distributed over $U(d)$ and
the difference between the two error curves seems to be
sub-exponential.

\section{Experimental realization}
\label{Experiments}

There are two different situations from the point of view of
experimental realizations. In many experiments several copies of a
quantum system are available at a time. Very often the systems
cannot be individually accessed. However, we would like that these
systems undergo {\it different} multilateral random unitary
rotations. In this case, according to our algorithm we have to
achieve that at each iteration step half of the systems undergo a
unitary rotation $U_k^{\otimes N},$ while the other half does not.
While numerically the mixing of the state $\rho$ and the rotated
one, $U_k^{\otimes N}\rho (U_k^{\otimes N})^\dagger$ is a matter of
adding two matrices, experimentally this mixing can be done with the
help of a controlled operation. The corresponding quantum circuit is
shown in Fig.~\ref{circuit}. For a single step of the algorithm we
realize $\twirl_2.$ The inputs are the state and an ancilla in a
superposition state, $(|0\rangle+|1\rangle)/\sqrt{2}$. The same
unitary, $U_k$ is applied on all qudits of the state, but only when
the ancilla qubit is in state 1. Finally, we measure the ancilla and
the outcome is $\tfrac{1}{2}[\rho + U_k^{\otimes N} \rho
(U_k^{\otimes N})^\dagger]$. This basic block is applied $M$ times,
each time using either a different, random unitary or a unitary from
the finite set as in Sec.~\ref{deterministic}. Note that the control
qubit can be a qubit which have a short coherence time compared to
the other qubits. It rapidly decays to state $\ket{0}$ or $\ket{1},$
and can be used as a sort of classical control for permitting the
unitary rotations on the other qudits.

In other experiments only a single copy of the state is produced.
For having an ensemble average of some quantity, the experiment must
be repeated many times. In this case our method can be used the
following way. Each time after the single copy of the state is
created, with $50\%$ probability we apply $U_1^{\otimes N},$ then
with $50\%$ probability we apply $U_2^{\otimes N},$ etc. The number
of gates needed in average is the half of the number of iterations.
If we use the deterministic version of our method described in
Sec.~\ref{deterministic} then it makes it possible to twirl with a
few single-qubit gates. This is an advantage in some systems. For
example, when using photons created with parametric down-conversion
and post-selection, the single-qubit gates can be realized with wave
plates.

\section{Numerical integration over $U(d)$}
\label{Numerics}

As we have already mentioned, our method can be used for integrating
numerically expressions of the type Eq.~(\ref{UUU}). In this section
we discuss how to generalize our approach for integrating
expressions of the type
\begin{eqnarray}
I:=&&\int_{U\in U(d)} \trace(A_1U) \trace(A_2U) ... \trace(A_m
U)\nonumber\\ &&\trace(B_1U^\dagger) \trace(B_2U^\dagger) ...
\trace(B_n U^\dagger) dU, \label{I}
\end{eqnarray}
where $A_k$ and $B_k$ are $d\times d$ matrices.

Based on the main ideas of the paper, Eq.~(\ref{I}) can be computed
in two steps. (i) First we need to obtain
\begin{equation}
M:=\int_{U\in U(d)} U^{\otimes m} \otimes (U^\dagger)^{\otimes n}
dU.
\end{equation}
This can be done by iterating the formula
\begin{equation}
M_{k+1}=\frac{1}{2} [ \openone_d^{\otimes (m+n)}+U_k^{\otimes
m}\otimes (U_k^\dagger)^{\otimes n}]M_k,
\end{equation}
where $M_0=\openone$ and $U_k$ are random unitaries. The series
$M_k$ will converge very fast to $M.$ (ii) The second step in
computing Eq.~(\ref{I}) is
\begin{equation}
I=\trace(M A_1\otimes A_2\otimes ... \otimes A_m \otimes B_1\otimes
B_2\otimes ... \otimes B_n).
\end{equation}
Note that $M$ does not depend on $A_k$ and $B_k.$ Thus when we
compute Eq.~(\ref{I}) for several $\{A_k\}$ and $\{B_k\}$, we have
to compute $M$ only once.

These ideas seem to work also when integrating over a subgroup of
$U(d),$ in particular, over the special unitary group $SU(d).$ Such
integrals appear, for example, in quantum
chromodynamics\cite{Urs,PP83}.

\section{Discussion}
\label{disc}

First let us discuss the importance of the fact that our algorithm
does not require an individual access to the systems of the
ensemble. This characteristic is important since we are presenting
the realization of a superoperator mapping a density matrix to
another density matrix. Ideally, we want that this mapping works
even if the density matrix describes an ensemble of very many
systems. A method which requires an individual access to the systems
of the ensemble cannot handle this situation. When realizing a
superoperator in a physical system, it is also advantageous that if
a pure state is mapped to a mixed one then this mixed state is
realized as the reduced state of a pure state of a larger system
\cite{HA04}. The usual method is not able to create such a
purification of the output density matrix. In contrast, our method
can handle a very large ensemble. Also, when we apply the quantum
circuit proposed in this paper for twirling, and we omit the
measurements then we get a pure state. The twirled state is the
reduced state of this pure state.

The algorithm presented in this paper is intimately related to other works
on random matrices. For instance, Ref.~\cite{PZ98} studies the statistical
properties of unitary matrices composed as the product of random
unitaries. In Ref.~\cite{EL05} it is proved that the product of a series
of random unitaries with nonuniform distribution converges exponentially
fast to the uniform distribution in many cases.

The relation of our paper to these works is the following. We also
used composed ensembles of unitary matrices. However, when looking
at Eq.~(\ref{twot}), we can see that our composed unitaries are not
independent and they are composed from a small set of random
matrices. Thus, especially in the first part of the paper, the main
goal was to realize twirling on an ensemble of {\it many} systems
using only a {\it few} random unitaries, rather than realizing
twirling using unitaries from an imperfect random source. Note that
we assumed that our unitaries were drawn from a perfect random
source providing unitaries distributed uniformly in $U(d).$ In the
second half of our paper we found that our results can also be
applied to the case of an imperfect source or for a deterministic
algorithm.

The key point of our algorithm is mixing of a subensemble in the
original state and the other subensemble in which all the systems
undergo the same multilateral rotation. This mixing can be realized
efficiently both in a classical computer and in a quantum computer.
Let us analyze the role of mixing pointing out something seemingly
paradoxical. Let us consider redefining $\twirl_2$ as the
application of a unitary $U$ which with $50\%$ probability it is the
identity and with $50\%$ probability it is uniformly distributed.
That would amount to applying a unitary with a distribution
\begin{equation}
  h(U) := \frac{1}{2}\delta(U-\openone) + \frac{1}{2}.
\end{equation}
However, if we compute the error of a single application of this
"new" $\twirl_2$ we find
\begin{eqnarray}
  \langle\Vert \hat{\rm \bf P}_2 \rho - \twirl \rho\Vert^2\rangle_h
  &=& \int \Vert U^{\otimes N}\rho(U^{\otimes N})^\dagger -
\twirl\rho\Vert^2 h(U)
  dU\nonumber\\
  &=& \int (\Vert \rho\Vert^2 - \Vert\twirl\rho\Vert^2) h(U) dU
  \nonumber\\
  &=& \Vert \rho\Vert^2 - \Vert\twirl\rho\Vert^2,
\end{eqnarray}
which unlike Eq.~(\ref{error_recursive_rho}), does not give us an
exponential convergence. Of course, this is because in the algorithm
described in this paragraph
 there is only one component present: The application
of a random unitary. The other component, namely
mixing of two subensembles,
is missing.

Moreover, while Ref.~\cite{EL05} studied the convergence of the
distribution of the composed unitaries to the uniform distribution,
we studied the convergence of a certain quantum operation, namely,
twirling. In particular, we computed the exponent of this
convergence and found that it does not depend on $N$ or $d.$ When
studying unitaries composed from random ones with a nonuniform
distribution, clearly the requirements for the convergence of the
operator built from these unitaries are much weaker than the
requirements for the convergence of the distribution of the
unitaries. It is also easier to get general statements on the
convergence of the operator for a quite wide class of faulty random
unitary generators. Indeed, we proved that convergence is reached
even in the case of a very poor random source.

Note that recursive algorithms can also be applied to summing over
discrete groups. For example, it has already been discussed in
Ref.~\cite{DL02} that a random element of the Clifford group can be
generated by a sequence of $O(n^8)$ operations. At each step, with
$1/2$ probability nothing happens, and with $1/2$ probability a
random element of the generating set is executed. Another example is
discussed in Refs.~\cite{AD05,KG05}. An $N$-qubit state can be
depolarized by summing over the stabilizer group \cite{G97,G98} of a
Greenberger-Horne-Zeilinger (GHZ,\cite{GH90}) state or a graph state
\cite{HE04,DA03} as
\begin{equation}
\rho=\sum_{k=1}^{2^N} S_k \rho_0 S_k^\dagger, \label{sumstab}
\end{equation}
where $\{S_k\}$ are the group elements. Exploiting that the
stabilizer group is commutative and that $S_k^2=1,$ the operation
Eq.~(\ref{sumstab}) can be realized in $N$ steps. At step $k$ with
$1/2$ probability nothing happens, and with $1/2$ probability $g_k$
is executed. Here $g_k$ are the $N$ generators of the stabilizer
group.

Finally, twirling a completely positive map \cite{DC06}, rather than
a quantum state, is also a useful procedure. Twirling a map makes it
possible, for example, to estimate the average fidelity of a
physical implementation of the map \cite{EA05}. It can be done with
a slight modification of our algorithm in three steps: (i) Applying
the circuit Fig.~\ref{circuit} several times with unitaries $U_1$,
$U_2$, $...$, $U_M$, (ii) applying the map, and (iii) applying again
the circuit Fig.~\ref{circuit} with unitaries $U_M^\dagger$,
$U_{M-1}^\dagger$, $...$, $U_1^\dagger.$ The control qubit for
rotations $U_k$ and $U_k^\dagger$ must be the same.

\section{Conclusions}
\label{conc}

Summing up, we have presented a very efficient approach for the
realization of twirling. Although it is based on random matrices, it
converges very fast, that is, the error decays exponentially with
the number of steps during the iteration. Together with the
simplicity of the method, this means that our algorithm requires
very little experimental or computational effort, for the
implementation in either a quantum or a classical computer. We have
demonstrated the robustness of the algorithm, which converges both
in the case of an imperfect random number generator and if the
unitaries are chosen deterministically from a small set. In the
future, it would be interesting to extend our approach to operations
which realize twirling with a subgroup of $U(d).$ In particular, we
would like to apply the method described in Section \ref{Numerics}
for integrating numerically over the $SU(d)$ group and look for
possible applications.

\acknowledgments

We thank M.M. Wolf for many useful discussions. We also thank P.
Hayden and U. Wenger for helpful comments. We acknowledge the
support of the EU projects RESQ and QUPRODIS and the
Kom\-pe\-tenz\-netz\-werk
Quan\-ten\-in\-for\-ma\-ti\-ons\-ver\-ar\-beitung der
Ba\-ye\-ri\-schen Staats\-re\-gie\-rung. J.J.G.R. is supported by
the Programa Ramon y Cajal of the Spanish Ministry of Education.
G.T. acknowledges the support of the European Union (Grant No.
MERG-CT-2005-029146), the National Research Fund of Hungary OTKA
under Contract No. T049234, and the J\'anos Bolyai Research
Scholarship of the Hungarian Academy of Sciences.

\appendix

\section {Computing the superoperators corresponding to twirling}
\label{SecSP}

For evaluating the right hand side of Eq.~(\ref{error_usualS}), we
need to know $\exs{\|S_{\twirl}\|^2}.$ For studying the convergence
through simulations of individual trajectories we need even the
matrix $S_{\twirl}.$ In this appendix, we will study the general
properties of $S_{\twirl}$ and determine it explicitly for small
systems.

Based on Eq.~(\ref{Sprop}) we can write
\begin{equation}
\exs{\|S_{\twirl}\|^2}=\trace(S_{\twirl}S_{\twirl}^\dagger)=
\trace(S_{\twirl}).
\end{equation}
Based on Eq.~(\ref{Sprop}) we also know that $S_{\twirl}$ is a
projector matrix with eigenvalues $0$ and $1,$ thus
$\exs{\|S_{\twirl}\|^2}$ must be an integer. We can get to know more
about $S_{\twirl}$ by recalling that twirling produces a Werner
state and such a state is a a linear combination of permutation
operators. Since these permutation matrices are not linearly
independent, one has to first orthogonalize them. Let us assume that
$\{R_k\}_{k=1}^{N_R}$ are the matrices obtained this way, satisfying
$\trace(R_kR_l)=\delta_{kl}$ where $\delta$ is the Kronecker symbol.
Now it is easy to see that we can write the twirled matrix as
\begin{eqnarray}
\twirl \rho=\sum_k \trace(\rho R_k)R_k.
\end{eqnarray}
Hence using Eq.~(\ref{vecA}), we obtain
\begin{eqnarray}
S_{\twirl}=\sum_k \vec{R}_k (\vec{R}_k)^\dagger.
\label{St}
\end{eqnarray}
Thus
\begin{eqnarray}
\exs{\|S_{\twirl}\|^2}=\trace(S_{\twirl})=N_R.
\end{eqnarray}

Now, let us determine $S_{\twirl}$ explicitly for two and three
qudits. For $N=2$ we have $N_R=2$ and a possible choice of the basis
matrices is \cite{W89}
\begin{eqnarray}
R_1&:=&\frac{\openone_d\otimes\openone_d+V_{12}}{\sqrt{d(d+1)}},\nonumber\\
R_2&:=&\frac{\openone_d\otimes\openone_d-V_{12}}{\sqrt{d(d-1)}}.
\end{eqnarray}
Here $V_{12}$ is the permutation matrix exchanging the two qudits
and $d$ is the dimension of the qudits.
Hence $S_{\twirl}$ can be reconstructed based on Eq.~(\ref{St}).
For $N=3$ and $d=2$ we have
$N_R=5,$ while for $d>2$ we have $N_R=6.$ The basis matrices can be
found in Refs.~\cite{EW01,EggelingPhD}.


\begin{thebibliography}{99}
\bibitem{AX05_cite} We note that replacing the integration by a sum
is not the only way to realize twirling. Twirling has been implemented in an
ensemble quantum computer (i.e., in an NMR quantum computer)
using a continuous set of rotations generated
by a field gradient over the sample \cite{AX05}.
\bibitem{otimes} We use the following definition for the tensor
product
\begin{equation}
A \otimes B=\left(
              \begin{array}{cccc}
                a_{11}B & a_{12}B & a_{13}B& ... \\
                a_{21}B & a_{22}B & a_{23}B& ... \\
                a_{31}B & a_{32}B & a_{33}B& ... \\
                ...     & ...     & ...    & ... \\
              \end{array}
            \right).
\end{equation}
\bibitem{norm} In this
paper we will present bounds for the Hilbert-Schmidt norm of the
error, i.e., the difference between twirling and our approximation.
Other norms can also be used for measuring the difference between
the $S$ matrices representing linear superoperators. The bounds we
present can be used to construct bounds also for these norms. In
particular, the trace norm is defined as
$\|A\|_{tr}:=\trace\sqrt{A^\dagger A}.$ Since for $n\times n$
matrices we have $\|A\|_{tr}^2\le n\|A\|^2,$ an upper bound for the
Hilbert-Schmidt norm of the error is also a valid bound for the
trace norm of the error. Instead of asking how large the difference
between the matrices representing two linear superoperators is, one
can look for the superoperator norm of the difference between the
two superoperators. Let $\mathcal{T}$ denote a mapping from a
density matrix to another density matrix. Then its superoperator
norm is defined as $\|\mathcal{T}\|_{so}:=\sup_{X\ne 0}
\|\mathcal{T}X\|_{tr}/ \|X\|_{tr}.$ For the norm of the difference
between two superoperators it can be proven that
$\|\mathcal{T}_1-\mathcal{T}_2\|_{so}^2\le
d^N\|S_{\mathcal{T}_1}-S_{\mathcal{T}_2}\|^2,$ where $\mathcal{T}$
acts on $N$ qudits of dimension $d.$ Here we used that for an
$n\times n$ matrix $A$ we have $\|A\|^2\le\|A\|_{tr}^2\le n\|A\|^2.$
Finally, another norm which can be used for measuring the difference
between superopertors is the diamond norm described in D. Aharonov,
A. Kitaev, and N. Nisan, quant-ph/9806029. The diamond norm is
defined as $\|\mathcal{T}\|_\diamond:=\|\mathcal{T}\otimes
(\openone_d^{\otimes N})\|_{so}.$ Here the tensor product by
$\openone$ represents the extension of the superoperator
$\mathcal{T}$ to a larger space. For our norm
$||S_{\mathcal{T}\otimes \left(\openone_d^{\otimes
N}\right)}||^2=d^{2N}||S_{\mathcal{T}}||^2.$ Hence it also follows
that $\|\mathcal{T}_1-\mathcal{T}_2\|_{\diamond}^2\le
d^{4N}\|S_{\mathcal{T}_1}-S_{\mathcal{T}_1}\|^2.$
\bibitem{SU} Clearly, it is enough if we are able to generate the
elements of the special unitary group $SU(2).$
\end{thebibliography}
\end{document}